# Large-scale neural recordings call for new insights to link brain and behavior

Anne E. Urai[1,2], Brent Doiron[3], Andrew M. Leifer[4] and Anne K. Churchland[1,5]

**1** Cold Spring Harbor Laboratory, Cold Spring Harbor, NY, USA;
**2** Leiden University, Leiden, The Netherlands; **3** University of Chicago, Chicago, IL, USA;
**4** Princeton University, Princeton, NJ, USA; **5** University of California Los Angeles, CA, USA
2021-07-23

## Abstract

Neuroscientists today can measure activity from more neurons than ever before, and are facing the challenge of connecting these brain-wide neural recordings to computation and behavior. Here, we first describe emerging tools and technologies being used to probe large-scale brain activity and new approaches to characterize behavior in the context of such measurements. We next highlight insights obtained from large-scale neural recordings in diverse model systems, and argue that some of these pose a challenge to traditional theoretical frameworks. Finally, we elaborate on existing modelling frameworks to interpret these data, and argue that interpreting brain-wide neural recordings calls for new theoretical approaches that may depend on the desired level of understanding at stake. These advances in both neural recordings and theory development will pave the way for critical advances in our understanding of the brain.

## Introduction

Our understanding of how the nervous system controls behavior closely trails our ability to precisely measure its core components - the activity of groups of neurons. Recent years have seen an explosion in large-scale neural recordings during animal behavior, opening up new ways to measure and understand network-level neural codes for cognition in diverse species. Here, we highlight how advances in technology have enabled this progress, reveal pitfalls and promises, and explain novel analysis approaches and theoretical tools being developed to understand the vast quantities of data now being collected. We focus on technologies for recording the activity of individual neurons. While measurements of neural mass signals (such as local field potentials, closely related scalp-level electrophysiological signals measured by EEG and MEG, widefield calcium imaging, fiber photometry, magnetic resonance imaging or functional ultrasound) have enabled advancements in studying large-scale brain networks, these techniques are beyond the scope of this review. We also exclude measurement of non-neural brain cells (such as glia) and extracellular signaling molecules (e.g. neuromodulators), as well as perturbation techniques such as optogenetics.

Each neuron does not act independently: the importance of not just studying neurons in isolation, but instead to understand simultaneous recordings of pairs of neurons, has been appreciated since the early days of neural recordings[1,2]. Some theoretical frameworks argued that correlations among neurons limited the information that a neural population could encode[3,4], while others emphasized the importance of the neural cell assembly, as a substrate for memory[5,6] and stimulus processing[7]. Consequently, there has been a concerted effort to understand the mechanistic origin[8] and computational role[9] of correlated fluctuations in neuronal population activity. The importance of understanding these interactions grows with the increasing sizes of simultaneously measured neural populations.

For decades, however, technical constraints limited many experiments to simultaneous recordings from one or at most a few cells. In practice, this meant that these neurons were usually hand-selected to respond strongly to experimenter-defined variables, such as visual motion or contrast. Many influential frameworks to understand neural computation then relied heavily on the neuron as a single unit, aiming to extrapolate or infer its role in local and long-range circuits[10–12]. In parallel, pioneering work in neuronal circuit models focused on capturing single neuron statistics such as firing rates[13] or spiking variability[14–16].

With the increasing throughput of simultaneous recordings in the '90s and '2000s (Box 1) came the (accurate) anticipation that large-scale recordings would speed up experiments and boost their statistical power. They also reduced the focus on hand-picked neurons and brain areas with well-characterized responses. What was less expected were the large changes in theoretical focus and a new depth of understanding. Big questions about how neural activity and behavior relate to each other are beginning to be within reach. For instance, how are neural representations distributed across brain areas and cell types? How do signals connected to task-related computations interact with signals related to other brain functions, such as movements and arousal? And how much of neural variability is truly stochastic 'noise', as opposed to a reflection of signals coming from other neurons, brain areas, or behaviors we couldn't measure before? Some skepticism is warranted as well: what have we learned from these advances in larger scale recordings and behavioral characterization, especially in small animals that allow for whole-brain recordings? Will large scale recordings deliver the promise of new insights, when many neurons in such recordings are unresponsive? And finally, where do current theoretical frameworks explain large-scale neural data, and where do they fall short?

Here, we aim to address these questions. We review how technological developments have brought increasing experimental throughput, allow large-scale surveys of neural responses across previously understudied brain areas, and are prompting new developments in studying information flow within and across brain areas. We then discuss how large-scale recordings have offered four unexpected insights (Figure 1):

- Neural representations of sensory/cognitive variables are distributed, sparse, and can be dwarfed by movement signals.
- Neural computations can be evident at the level of population dynamics but hidden at the level of single neuron firing rates.
- Behaviorally relevant neural variance can often be explained by a small number of dimensions.
- Largely unstructured network architectures can drive highly structured responses.

We elaborate on existing theoretical frameworks for interpreting neural data, and how they are challenged by results from large-scale recordings. Finally, we speculate about future experimental and theoretical developments, and explore skepticism regarding the role of large-scale recording in helping us understand brain function.

# Insights from large-scale neural recordings

Recent studies have leveraged high-yield recording modalities (Box 2) in the hopes of gaining new insights into behavior and brain activity. The ability to record many neurons at once has increased statistical power, and reduced the number of required research animals. It has also shifted the focus from hand-selected neurons and brain areas toward unbiased, global surveys of neural responses and cell types. Large-scale recording techniques also increase the probability of encountering neurons with rare responses, capturing small signals that are distributed over many cells, or recording rare cell types. For instance, matching high-density electrophysiological responses to anatomically identified cells makes it possible to define the responses of sparse cell types in the retina with distinctive morphology[17]. Furthermore, leveraging the simultaneous nature of large-scale neural recordings has brought new insights, by shifting focus from single units to larger neural populations and brain regions.

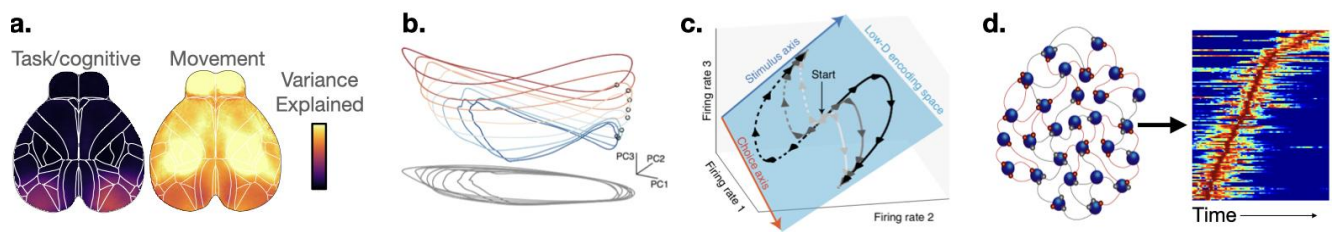

*Figure 1. Insights from large-scale neural recordings. (a)* Neural representations of task/cognitive variables (left) are distributed, sparse, and can be dwarfed by movement signals (right). Adapted from[18]. *(b)* Neural computations can be evident at the level of population dynamics but hidden at the level of single neuron firing rates. Adapted from[19]; colors show neural responses in motor cortex during cycling at different speeds; each loop is once around a repeating cycle; blue is slowest. *(c)* Behaviorally relevant neural variance can often be explained by a small number of dimensions (blue, red axes). Adapted from[20]. *(d)* Largely unstructured network architectures (left) can drive highly structured responses (right) in neurons. Adapted from [21].

Large-scale recording efforts in rodents have begun to generate new insights into the **distributed and sparse nature of neural responses to task-relevant information** (Figure 1a). In one recent study, Steinmetz et al.[22] used Neuropixels probes to measure the responses of ~30,000 neurons in mice reporting spatial judgments about visual stimuli. Choice signals were sparse, widely distributed, and plentiful in deep structures that had thus far been overlooked in studies of decision-making (e.g., the midbrain reticular nucleus). Interestingly, only a small fraction (~18%) of neurons in V1 were responsive to visual grating stimuli. By showing mice a larger battery of visual stimuli (including drifting gratings, Gabors and flashes), another Neuropixels study[23] could identify receptive fields in up to 70% of V1 neurons. The recording method matters here: since spike-sorted electrophysiology is biased towards cells with large or frequent spikes, it is more likely to overestimate the fraction of neurons responsive to visual stimulation[24,25]. Indeed, when de Vries et al.[26] used calcium imaging to measure the activity of ~60,000 neurons in passively viewing mice (viewing the same set of stimuli as used in [23]), they found that few neurons were driven by static, abstract stimuli, and many neurons were not responsive to visual stimuli at all. While the activity of some neurons could be well predicted by their response to visual stimuli, many more cells responded in ways that could not be captured by existing models of cortical function. While the presence of so-called 'dark neurons' (neurons that don't fire at all[27], or don't fire in responses to experimental variables[25]) has been long known, results from large-scale recordings are emphasizing the importance of understanding their prevalence and functional role[25].

Large scale recordings also showed that **neural activity correlated with animal movements, including idiosyncratic, task-unrelated "fidgets," is stronger and more widely distributed than previously thought**[18,28,29](Figure 1a). While it has long been known that running modulates activity in visual cortex[30–33], recent work extended these findings in important ways. First, the impact of movements on single neurons was not restricted to a single area[18,28,29]. Second, movement-driven modulation was present even in animals who were not just passively viewing, but were instead engaged in expert cognitive behaviors[18,29]. While the existence of movement-related activity in untrained, passively viewing animals is perhaps unsurprising (after all, not much else is being asked of the brain in such circumstances), seeing movements dominate even in expert, engaged animals was unexpected: one would assume that task-related signals would dominate. Third, by using unsupervised video analysis, recent studies considered a far greater diversity of movements beyond running and pupil diameter[18,28], and evaluated these movements' role in neural activity in a hypothesis-free way. This confirmed the importance of well-known movements such as pupil dilation, and also revealed previously ignored movements, such as hindlimb flexions and orofacial movements. The strong effect of facial movements on neural activity may indicate that these movements reflect animals' emotional state[34]. An additional reason that orofacial and other movements are so widespread, may be that many brain areas need to predict the sensory consequences of impending movements[35].

An anticipated challenge of the large scale recordings described above was the need to analyze them. What was not anticipated was that the increased neuron count in large scale recordings would reveal that **seemingly complex dynamics often reflect the sum of a small number of underlying motifs**[36,37] (Fig. 2c). Specifically, neural population recordings have uncovered that variance in neural activity, and specifically variance relevant for behavior, can often be accounted for by a small number of dimensions (Figure 1c). For instance, a choice decoder built on the first principal component of neural activity in monkey higher visual cortex performed almost as well as one built on the whole dataset[38]. Low dimensional activity is also apparent in frontal cortex of monkeys performing complex tasks[20], in monkey premotor cortex during reaching[39] and in *C. elegans* during fictive locomotion[40]. In another example[41], researchers recorded the activity of ~150 neurons from the isolated nervous system of a medicinal leech. They used dimensionality reduction to identify an axis in neural state space, along which the population's activity predicted behavioral responses to sensory stimulation (swimming or crawling) earlier in time than any single neuron. From the cells that strongly contributed to this population, they also identified one specific cell that could bias decisions towards crawling when electrically stimulated.

To what extent are neural codes truly low-dimensional? Theoretical work suggests that multidimensional representations are critical[42,43] and that the limits of dimensionality reduction must be noted[44]. In support of this, recent experimental work[45] showed that the dimensional structure of neural responses in V1 allows for a code that balances efficiency with robustness to small perturbations in visual images.

Although the observations above benefited from large-scale recording technologies, some may in principle have been made with many neurons recorded sequentially. Simultaneous recordings additionally uncover the relationships among neurons, such as the way that their responses change together. For instance, internal states such as engagement tend to drive large fluctuations that are shared between many neurons: this variability would have looked like random trial-to-trial noise when recording one neuron at a time. Simultaneous large-scale neural recordings have enabled several important insights (below) that have changed the way we analyse and think about neural population activity.

Results from simultaneous recordings challenge classical population coding approaches which focus on distributed input tuning over a set of neurons[46,47]. This view treats the population as simply a collection of individual neurons, where decoders estimate inputs using a suitable weighting of static neuronal responses. Such population codes work well when a majority of neurons show a simple and straightforward tuned response. However, large-scale recordings are **revealing neural computations (such as movement planning and decision making) that are evident at the level of population dynamics, even when single neurons do not show an obvious tuning to stimulus or task variables**[19,22,48,49]. Such heterogeneous activity has prompted new frameworks in which the representation is contained in the dynamics of the population response[50] (Figure 1b).

Another major insight that has been gained from simultaneous recordings is that **transient or fluctuating responses at the single neuron level can give way to stability at the population level** (Figure 1d). For example, persistent activity in single neurons was once thought to be the sole substrate of slow-timescale cognitive processes, such as working memory[51,52] and were traditionally modelled by fixed point attractors[15]. Recent work extends this idea by demonstrating that dynamic single neuron responses can coexist with a stable, lower dimensional, subspace coding that offers comparable benefits[53]. Moreover, populations of transiently responding neurons can be generated by networks with minimal structure, and offer benefits over persistent activity in single neurons in terms of robustness and flexibility [21].

Large scale simultaneous recordings have also **challenged the role of sequentially firing neurons** (Figure 1d). The predominant theory was that sequences of neurons reflect highly structured neural circuits, such as synfire chains[54]. However, large scale recordings uncovered that instead, sequential firing can emerge through cooperation between recurrent synaptic interactions and external inputs[21], which argues that neural sequences can emerge gradually from largely unstructured network architectures. These networks offer benefits over persistent activity in single neurons in terms of robustness and flexibility. Hippocampal replay, thought to be involved in memory consolidation, offers another example of sequential activation that can be best studied when populations of many neurons are recorded simultaneously[55]. Observing a hippocampal replay event requires precisely noting the relative timing of a population of place cells on the tens of milliseconds timescale. The insight that hippocampal activity replays an animal's previous experience, forwards or backwards in time, thus relies crucially on the ability to observe relative timing among simultaneously recorded neurons.

A final advantage of simultaneously recorded neurons is that co-fluctuations among neurons offers insight into multi-region communication[56]. For instance, some slow drifts in neural responses are shared between V4 and PFC and can predict a monkey's fluctuating impulsivity in a decision-making task over the course of an experimental session[57]. Because the direction of the drifting signal was diverse across neurons, it could only be uncovered by analysing large-scale simultaneous recordings. Other analyses of communication across cortical areas have revealed that some activity fluctuations are communicated to downstream structures, while others remain private[58]. Such approaches are crucial for understanding ever-larger, multi-region neural recordings and promise a shift towards understanding neural activity in brain-wide models.

Analyses of multi-region communication often infer connectivity via correlations[59] in part because the connections of each neuron within an area are not known. In animals for which full[60,61], or partial[62,63] connectomes are known, interpreting multi-region recordings is becoming more concrete. In drosophila, neural recordings had previously identified neurons that track the animal's heading direction, putatively functioning as a ring-attractor network[64,65]. However, the connectome made it possible to extend these physiological results to arrive at a circuit model for how these neurons, and their connection weights, can compute a transformation from egocentric to allocentric coordinates. The critical observation from the connectome was the precise offset of synaptic weights between two cell-types (PFN and hΔB)[66–68]. This approach will likely benefit rodent researchers likewise seeking to understand how allocentric travelling direction is computed[69]. New experiments can evaluate whether the concrete model predictions garnered from drosophila[68] are realized in the rodent.

# Box 1: History and future of large-scale neural recordings

Can we ever expect to record all neurons in the brain simultaneously? The answer depends on the size and physical properties of the brain in question. Simultaneous whole-brain measurements of single-neuron activity have been acquired in small, transparent animals, notably C. elegans [70], larval zebrafish [71], hydra [72] and perhaps soon Drosophila [73,74]. In mammals, electrophysiological recordings across all cortical neurons has not been achieved but may be possible in principle [75,76].

In 2011, Stevenson and Kording [77] proposed a 'Moore's law' for neural recordings, predicting a doubling of simultaneously recorded neurons every +- 7.4 years. This prediction has been borne out, and calcium imaging methods have seen even faster increases than predicted. Yield from imaging, however, comes with trade-offs in temporal resolution, signal-to-noise ratio or imaging depth.

Exponential increases in recording ability offers exciting prospects for whole-brain imaging in larger brains, but we remain orders of magnitudes away from recording a sizable fraction of the mammalian brain. Extrapolating suggests that whole-brain, single-neuron recordings in mice may become reality between two decades and a century from now.

Are whole-brain recordings necessary to understand all aspects of nervous system function? Whole-brain recordings in small animals may hold valuable lessons: in systems such as C. elegans, hydra and zebrafish, whole-brain recordings could be subsampled to assess the value of measuring a more and more complete set of nervous system activity.

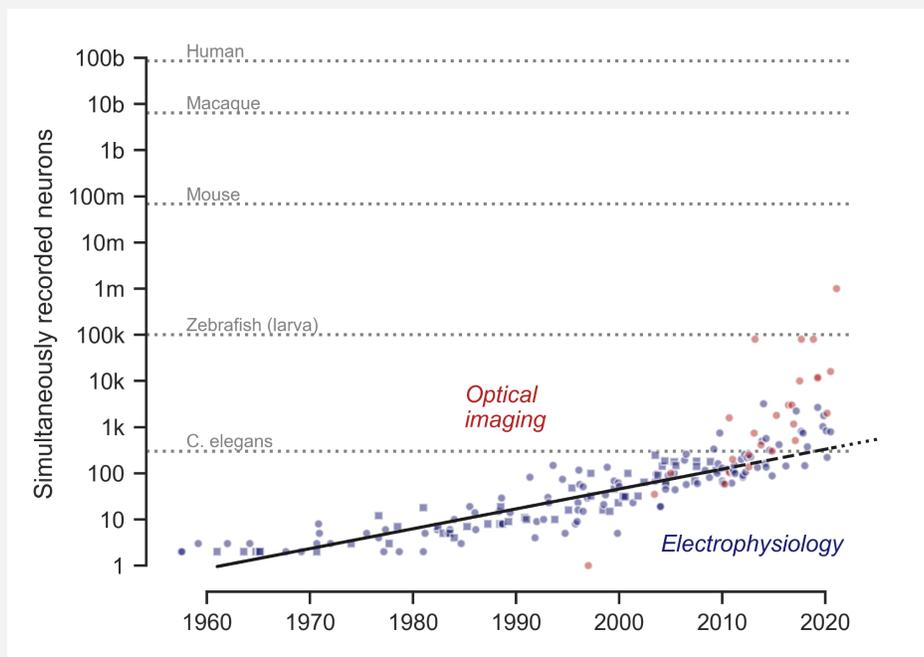

*Figure 2. Scaling up of neural recordings.* Dark blue points: number of simultaneously recorded neurons using electrophysiology (squares indicate the original data used for the fit (Stevenson and Kording, 2011, their Appendix 1)). Red points: measurements using optical imaging (two-photon or light-sheet). Line: exponential fit to the original data. Black dashed/dotted lines: extrapolation to the present/future. Gray lines: approximate number of neurons in the brains of commonly used species in neuroscience. Figure available at https://github.com/anne-urai/largescale_recordings under a CC-BY license.

## Box 2. Tools and technologies to observe brain and behavior

Recent decades have seen dramatic progress in the ability to record and process neural activity, and to connect it to behavior. The number of simultaneously, electrically recorded cells has been increasing, and the development of Neuropixels probes accelerated this considerably. These linear probes with up to ~10,000 recording sites allow simultaneous recording of large neural populations spanning multiple areas [23,78,79] (Figure 3c). Further, the thin shank (70 × 20 μm) causes minimal tissue displacement relative to previously used braided wires, and the probe's active circuits for amplifying, digitizing and multiplexing lowers noise levels. Improvements in flexible electrodes, either part of an injectable mesh or inserted by a stiff guide, may further help to reduce tissue damage, and can enable stable, long-term interrogation of neural circuits during behavior [80,81].

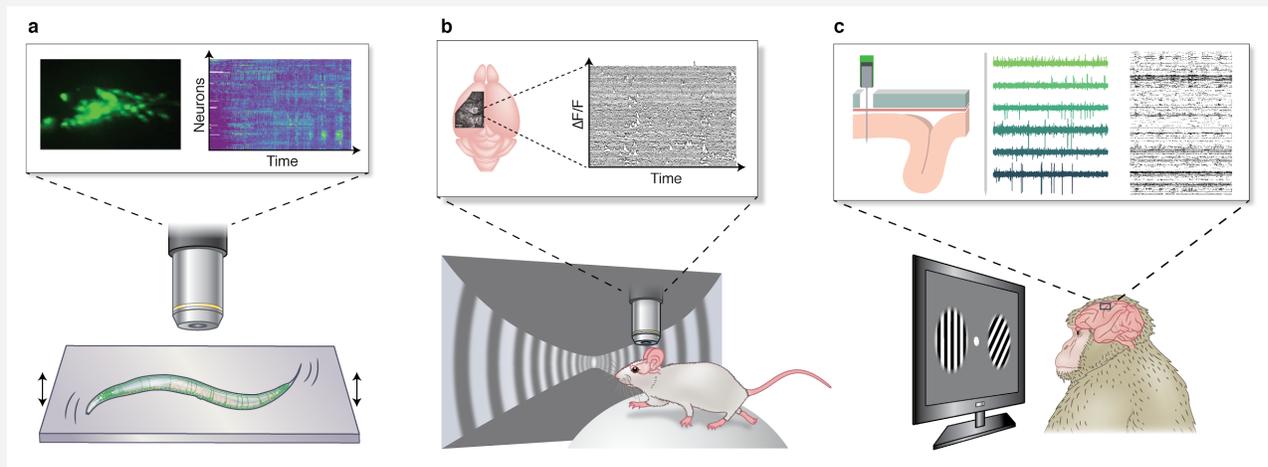

*Figure 3. Large-scale neural recordings in behaving animals. (a) Confocal microscopy of all neurons in C. elegans as it freely moves on an adjustable platform. Adapted from [82] under a CC-BY license. (b) Mesoscope 2-photon imaging of the cortical surface while a mouse moves through a virtual reality by running on a ball. Adapted from [83] under a CC-BY license. (c) High-density electrophysiological recordings using Neuropixels probes, while a monkey performs a psychophysical decision-making task.*

A major challenge for large-scale electrophysiology is a lack of consensus on spike sorting. While recording from densely spaced sites can facilitate automated spike sorting, the problem still requires significant manual curation [84]. Certain experiments may tolerate imperfect spike sorting [85], but many demand confidence in knowing which neuron was recorded. Simultaneous juxtacellular and extracellular recordings can provide "ground truth" benchmark data on spike sorting, against which different algorithms have been systematically evaluated [86]. These approaches are moving the field forward from traditional laboratory-specific and manually curated spike sorting towards standardized, less subjective practices [87].

Optical imaging has recently overtaken electrophysiology in its neural yield (Box 1). Recent progress has pushed the number of simultaneously recorded cells in the mouse brain to one million, about a tenth of its cortex [88]. Imaging has high spatial resolution and coverage, allows for labeling of specific cell types or projection targets [89], and can reveal the spatial organization of activity patterns [64]. Imaging through thinned skulls, cranial windows or in transparent animals is minimally invasive, and allows for long-term monitoring of the same structures (Figure 3b). Head-mounted microscopes additionally allow imaging in freely moving animals, and targeting of deep structures [90,91]. However, since both calcium kinetics and the dynamics of calcium indicators are slow compared to neural firing, calcium imaging can only be used as a coarse proxy for spike timing and rate [24,92,93].

In smaller, transparent animals such as larval zebrafish, *C. elegans*, and hydra, activity of the majority of neurons can be imaged simultaneously at cellular resolution in the fully intact animal [70–72]. *C. elegans* recordings, for example, have provided a sandbox to test new theories of non-linear dynamical systems models applied to whole brain dynamics [94–97]. Fast-tracking microscopes allow for large scale population imaging even as the animal swims or crawls [82,98–101] (Figure 3a) and have revealed the importance of population codes for representing locomotion even in relatively simple animals [102]. Multicolor labeling strategies to register entire brains onto an atlas with single-neuron accuracy [103] now allow whole-brain activity to be compared across individuals at cellular resolution and to further be linked to gene expression [104], opening up new ways to study individual variability in neural coding.

A final technological advance is in new ways of quantifying behavior more fully, largely driven by progress in video tracking and processing [105–107]. Such data-driven approaches in parsing spontaneous behavior [108–111] can allow us to interpret neural activity in the context of the behavior it produces, with or without experimenter-imposed task structure. This has benefitted both traditional, well-controlled behaviors [112,113] as well as more ethological ones [114–116].

# Theoretical frameworks: more is different

A potential criticism of large-scale recording studies is that they are observational rather than hypothesis-driven, and lack the ability to distinguish concrete mechanistic models. In some sense, this is a fair criticism. A full understanding of brain function will require more than simply a list of all neurons and the extent to which each is modulated by one variable or another. On the other hand, some current studies deploy descriptive models partly out of necessity, as the complexity of the measured activity can make it difficult to relate it to existing theoretical frameworks. In the past, optimism suggested that if only we could record more of the right neurons, our elegant *a priori* models would be readily confirmed. But when one records neural activity, the diversity of variables that modulate neural activity makes it challenging to argue that a signal at hand truly reflects a hypothesized computation. For instance, models of evidence accumulation offer appealing explanations for decisions made in the face of noisy evidence [117,118]. Neural activity that "ramps" during decision formation is certainly reminiscent of evidence accumulation [119], but such ramps can also reflect idiosyncratic combinations of stimuli and movements [18] or the average of multiple, disparate sensory and decision-related motifs [120]. Thus, large scale recordings are currently uncovering such unexpected neural responses that the ability to connect them to theoretical mechanisms may seem, at least momentarily, out of reach. A likely way forward is that the current focus on detailed characterization of neurons across brain structures will give way to more hypothesis-driven experiments in the near future. In this section we discuss how new datasets can start to inform physiological models of large scale brain networks, and point to the need for new (and different) theoretical frameworks to integrate brain and behavior.

What is needed is a blueprint for how large-scale datasets can first inform and then be a test for mechanistic models of cortical circuits. One promising approach to creating such models is to better understand the causes and consequences of variability in neural activity: trial-to-trial neuronal responses and within-trial spiking dynamics in diverse brain areas are famously variable[121] (Figure 4a-c). Physicists have long used variability as a window into the dynamical interactions between components of a larger complex system[122]. In the brain, any response variability reflects the underlying biology of the nervous system; however, the vast spatial and temporal scales over which this biology operates makes it challenging to unravel the underlying neuronal mechanics. Indeed, past modelling efforts have been constrained by the experimental techniques used to record from local, small scale cortical circuits[14,16,123–126]. A severe limitation of this approach is that these models must make assumptions about any variability inherited from outside the circuit[127].

We here review how large-scale recordings can alleviate this shortcoming, and can contribute to understanding neural variability across spatial scales. Rather than aim for a single model of the brain (perhaps a large-scale simulation that gives rise to the sorts of computations observed in real brains), we include models of varying complexity; this is a hybrid (rather than hierarchical) approach that allows high-level and fine-grained models to coexist, each explaining different features of the data [128].

Perhaps best categorized is variability at the smallest scales of membrane and synaptic dynamics (Figure 4d). Here, seminal work has shown how synaptic vesicle release and recovery is very unreliable[129,130] and voltage-gated ion channels in the cellular membrane open and close randomly [131,132]. At a larger spatial scale, neuronal recordings have provided significant evidence of the spiking variability of single neurons, and populations of neurons in a local circuit (Figure 4e). This variability is so pervasive that successful statistical modeling frameworks often take neurons to behave like *a priori* Poisson processes[133,134]. One often cited mechanism underlying such spiking variability is the emergent population dynamics in networks with strong and balanced excitatory and inhibitory recurrent interactions[14,15]. Recent advances in balanced excitatory-inhibitory networks with structured connectivity account for correlated pairwise variability[125,135,136], and even low dimensional population-wide shared variability[127,137,138]. Extending such mechanistic understanding to how variability is distributed over multiple brain regions remains a significant challenge in the new era of large datasets.

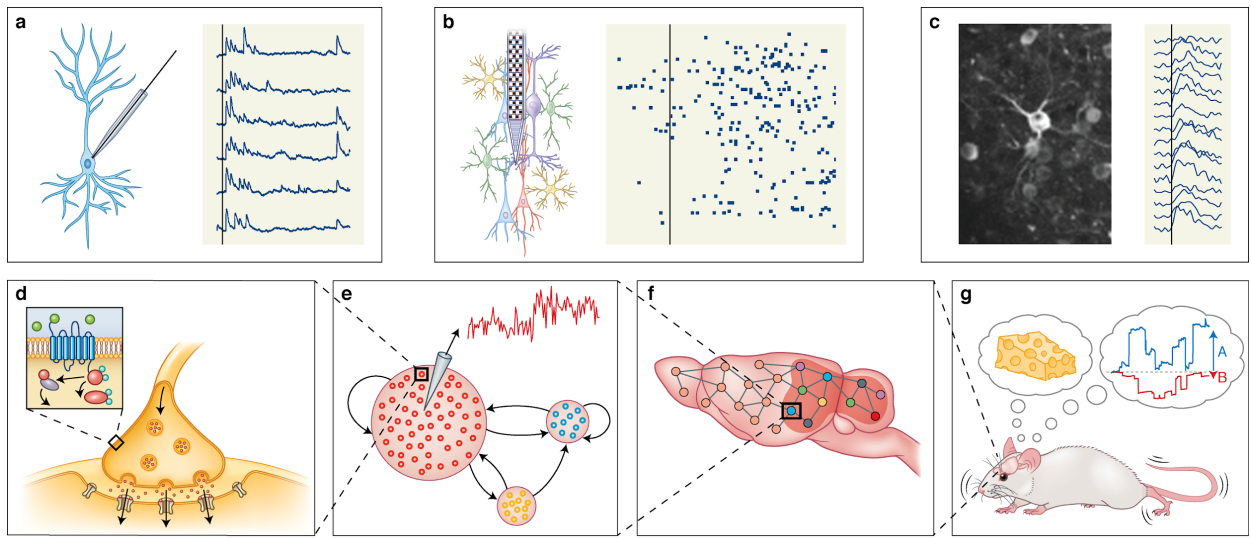

*Figure 4. Understanding trial-to-trial neural variability across scales. (a-c) Trial-to-trial variability of single-neuron responses in different recording modalities. (a) Variability in responses to pre-synaptic stimulation, measured using patch clamp in vitro. Adapted from [121]. (b) Variability in spike timing and rate to visual stimulus presentation, measured using extracellular silicon probes in mouse visual cortex. (c) Variability in calcium responses to whisker stimulation in mouse somatosensory cortex. Adapted from [139] with permission from the author. (d-e) Understanding neural variability across spatial scales. (d) Cellular noise at the level of synapses and membrane dynamics. (e) Circuit noise, arising from the dynamics of local populations of excitatory and inhibitory neurons. Adapted from [140] under a CC-BY license. (f) Whole-brain noise, arising from the interactions between brain areas that may propagate or dampen variability. (g) Interpreting neural variability in the context of animal behavior, quantified from e.g. computational models of task-related cognitive processes, body movements and pupil-linked arousal, and task-unrelated physiological states.*

Compared to cellular or local network scales, much less is known about how the trial-to-trial neural variability in one brain region depends upon the neuronal variability distributed over the rest of the brain (Figure 4f). Large-scale recordings are starting to show how spiking variability in one brain region is inherited or filtered by another, prominently along the visual pathway[56,141], and in the songbird system[142]. Yet it is not clear if all the variability in one region should be attributed to outside sources[143], or that some component is internally generated from interactions within the circuit itself[123,127,144]. Promising new analysis methods have started to identify specific activity patterns in one brain area, whose variability is inherited from activity in an upstream region: a 'communication subspace' [58]. In the macaque visual system, only a small subset of the scope of V1 population variability drives population variability in V2, and this V1-V2 predictive dimension is largely non-overlapping with the feedback V2-V1 subspace [58]. This may allow V1 to route selective activity to different downstream areas and reduce unwanted co-fluctuations in downstream areas [145]. Such dimensionality reduction approaches to link variability across brain regions will be needed to keep any ensuing mechanistic models tractable in the era of large-scale neural recordings [146].

Recent work has increasingly appreciated the large role of internal states [147,148] and rich, spontaneous behaviors [18,28,109] as crucial predictors of trial-to-trial neural variability (Figure 4g). As these studies have remained largely descriptive, future theoretical approaches are needed to integrate the presence of such varied signals with the core computations carried out by neural circuits. As neural recordings increasingly capture most or all of the brain (Box 1), distinct implications for theoretical models may arise. In one extreme, ever-larger neural population recordings may reduce the need for detailed behavioral quantification for capturing neuronal variability. In this view, behavior is simply a proxy for (yet to be measured) neuronal activity. Alternatively, behavior may be a complex expression of distributed neuronal activity and must be an equal partner in any comprehensive model of brain activity and its variability. Ultimately, a resolution to this issue may lead us to fully understand the brain as part of the whole animal, with an appreciation for its evolutionary past and ethological niche [115,149].

The possibility of whole-brain recordings in small animals shows us both promise and warning: larger observations bring more nuance, but also lay bare gaps in our tools for interpreting brain-wide neural dynamics. Large scale recordings in *C. elegans*, zebrafish and *drosophila* have revealed that many neurons are tuned to diverse aspects of behavior, often in subtle ways [102,150–153]. These small systems may be ideal models in which there is a semblance of 'ground truth' for testing new theoretical frameworks, before applying them to larger brains [154]. While they provide striking opportunities for demonstrating the predictive power of such recordings, for example by 'mindreading' animal behavior, they also remind us of how much of neural activity (measured by variance explained or otherwise) we still don't understand.

## Conclusions and outlook

Over the last few decades, our ability to perform large-scale neural recordings during behavior has grown by orders of magnitude: a postdoc can now record more neurons in a day than her PI was able to collect over the course of an entire postdoctoral fellowship. We here reviewed the technical progress and major insights gained from such experiments. We have highlighted how these advances have answered key questions in the field, and raised new ones for which theoretical frameworks are only starting to be developed.

With the progression of large-scale recording technology and computational capacity, what can we expect in the years and decades to come? Accurately predicting and decoding behavior from neural activity will likely become available for species with larger brains, building on the successes of small invertebrate models. Especially for brain-computer interfaces, this may have a significant impact on translational neuroscience. More simultaneously recorded neurons across connected brain regions will give a tighter handle on the sources of neural variability, and the distributed nature of neural circuit computations. We also expect to see an increasing appreciation for how neural computations depend on internal states, individual animals' individual life history and a diversity in behavioral strategies. We hope that in future work, deep understanding of animal behavior (from psychophysics to body movements and ethology) will be central to interpreting neural data. Improvements in dimensionality reduction, going beyond linear techniques such as principal component or factor analysis, will be needed to answer important questions about the size and complexity of neural circuits required for specific computations. More direct cross-species comparison will facilitate the transfer of insights from smaller, more tractable brains to larger organisms such as ourselves.

Ultimately, just recording many neurons - even when accompanied by well-quantified behavior and a connectome - will be insufficient to fully understand the brain: a full, directed, network description with all synaptic weights may be another core missing component for performing causal inference and behavioral prediction. The completeness of whole-brain recordings also stands in contrast to the inaccessibility of other signals in the brain, including neuromodulators [155], glia and glia-like cells [156]. We also don't yet fully understand how neural recordings we observe relate to neural wiring [154] or gene expression [104]. Some argue that even in a world with perfect and complete data, we may not be able to understand the brain to our satisfaction. This question has recently led to lively debates, often centering around the fundamental question of what constitutes an explanation in neuroscience. Given the immense progress of the past few decades, we optimistically predict that the future will continue to bring in-depth, unexpected and multifaceted understanding of brain function in all its complexities.

## Data and code availability

All code and data used to generate Figure 2 is available at
https://github.com/anne-urai/largescale_recordings under a CC-BY 4.0 license.

## Acknowledgements

A.E.U. is supported by the German National Academy of Sciences Leopoldina and the International Brain Research Organization. B.D. is supported by NIH 1U19NS107613-01, R01EB026953; Vannevar Bush faculty fellowship N00014-18-1-2002 and the Simons Foundation Collaboration on the Global Brain. A.M.L. is supported by the National Institute of Neurological Disorders and Stroke of the National Institutes of Health under New Innovator Award No. DP2NS116768, and Simons Foundation Award #SCGB #543003. A.K.C. is supported by NIH R01EY022979, NIH R01EB026949 and the Simons Collaboration on the Global Brain.
We thank Nicholas Sofroniew for sharing the mesoscope image panel shown in Figure 3b, Eric Trautman and Krishna Shenoy for the primate electrophysiology (Neuropixels) data in Figure 3c, and Debbie Maizels for graphic design. Ian Stevenson, Karel Svoboda, Peter Rupprecht, Adam Charles and Guido Meijer suggested data points shown in Box 1, and João Couto provided helpful comments on an earlier version of the manuscript. John Tuthill provided insights on interpreting data from the fly connectome.

## Citation diversity statement

Recent work in several fields of science has identified a bias in citation practices such that papers from women and other minority scholars are under-cited relative to the number of such papers in the field [157]. Here we sought to proactively consider choosing references that reflect the diversity of the field.
The gender balance of citations was quantified from the first names of the first and last authors [158]. Expected proportions estimated from five top neuroscience journals since 1997 are 6.7% woman/woman, 9.4% man/woman, 25.5% woman/man, and 58.4% man/man [157]. By this measure, our references (excluding those before 1997) contain 9.5% woman/woman, 10.9% man/woman, 21.9% woman/man, and 57.7% man/man.